\documentclass[twocolumn,showpacs, preprintnumbers,amsmath,amssymb]{revtex4}

\usepackage{graphicx}
\usepackage{amsmath}

\begin{document}

\title{Thermodynamic coherence of the Variational Average-Atom in Quantum Plasmas (VAAQP) approach.}


\author{R. Piron$^1$\footnote{Electronic address: robin.piron@cea.fr}, T. Blenski$^1$ and B. Cichocki$^2$}
\affiliation{$^1$ CEA, IRAMIS, Service des Photons Atomes et Mol\'ecules, F-91191 Gif-sur-Yvette, France.\\
$^2$ Institute of Theoretical Physics, Warsaw University, Ho\.{z}a 69, PL 00-681 Warsaw, Poland.}


\date{\today}

\begin{abstract}
A new code called VAAQP (Variational Average-Atom in Quantum Plasmas) is reported. The model as well as main results of previous studies are briefly recalled. The code is based on a new fully variational model of dense plasmas at equilibrium with quantum treatment of all electrons. The code can calculate the Average Atom structure and the mean ionization from the variational equations respecting the virial theorem and without imposing the neutrality of the Wigner-Seitz sphere. The formula obtained for the electronic pressure is simple and does not require any numerical differentiation. A description of the principal features of the code is given. The thermodynamic consistency of the results obtained with VAAQP is shown by a comparison with another approach on the example of the aluminium $10$ $eV$ isotherm EOS curve. A first comparison to an \textsc{Inferno}-type model is also presented.
\end{abstract}



\maketitle

\newcommand{\ea}{{\it et al.~}}
\renewcommand{\rho}{\varrho}
\renewcommand{\phi}{\varphi}

\section{Introduction}
Models of plasma at local thermodynamic equilibrium taking into account bound electrons are being developed since long time ago. Such models first appeared in astrophysics in order to calculate equation of state (EOS) and radiative properties of stellar objects. Calculations taking into account atomic electrons, allowing one to estimate the effective ion charge and photon absorption cross sections are essential for the modeling of stars. The ability to account for bound electrons is important since it is well known that even relatively low $Z$ elements are not totally ionized at temperatures typical for the central parts of the sun. More recently plasma atomic models taking into account bound electrons have also become important in the inertial confinement fusion (ICF) research, especially for the hydrodynamics modeling of ICF targets. The role of the atomic physics models and codes in this modeling is to provide plasma EOS as well as transport coefficients like opacities and resistivity. With the development of the laboratory dense plasmas such models are also useful in plasma spectroscopy. Among the first models of atoms in plasmas was the well-known Thomas-Fermi (TF) spherical ion-in-cell model proposed by Feynman \ea \cite{Feynman49}. This model based initially on ideas from the solid state physics extended to finite temperatures appeared to be thermodynamically consistent and successful in many practical applications. It is based on a variational principle, respects the virial theorem and provides a relatively simple expression for the electronic pressure. The plasma equilibrium can be constructed for each values of the three parameters : atomic number $Z$, ion density $n_{i}$ and temperature $T$. The approach of Ref.~\cite{Feynman49} is to model the whole bulk plasma through a self-consistent calculation of a chosen central average atom treated within the quasi-classical TF statistical model. Non-central ions and electrons of the plasma may be viewed as a homogeneous jellium existing at the border and beyond the boundary of the central ion spherical volume which is electrically neutral. The radius of this sphere is the Wigner-Seitz radius $R_{WS}$ related to the ion density by the formula $R_{WS}=\left(3/(4\pi n_{i})\right)^{1/3}$. The electron density of the jellium $n_{0}$ related to the number of free electron per atom $Z^{*}=n_{0}/n_{i}$ is not known a priori. It is found from the ion-in-cell TF model self-consistent field (SCF) calculation and equals to the electron density at the radius of the WS sphere. Similarly the electronic pressure in this model appears to be the pressure of the homogeneous electron gas at the atomic sphere boundary. The TF EOS is still being used in many ICF and astrophysical applications (see for example \cite{More88}). Its quasi-classical character does not allows it to be used for the modeling of the radiative properties.

For this reason with the appearance and progress in computers there have been several attempts towards quantum extensions of the average atom model. The most widely used and successful model has been proposed in 1972 by Rozsnyai (see \cite{Rozsnyai72}). In Rozsnyai's approach the free electrons have still been treated using the TF approximation. This is the principal reason why this model is not based on a variational approach and does not respect the virial theorem. The first entirely quantum model has been the \textsc{Inferno} model proposed in 1979 by Liberman (see \cite{Liberman79,Wilson06}) in which both bound and free electrons were treated quantum-mechanically. Liberman has derived the self-consistent-field equations for the electrons from a variational principle however the average ionization in his model is obtained from the neutrality of the WS sphere assumed in analogy to the TF ion-in-cell model. Both bound and free electron wave functions in Liberman's model are normalized in the whole space. Liberman's model does not respect the virial theorem and does not have a simple expression for the thermodynamic pressure. Instead the pressure has to be calculated via a numerical derivation of the free energy per ion with respect to the ion density. One of the essential points in this model is the neglect of the Friedel oscillations \cite{Khanna76,Gouedard78} beyond the WS radius.

These oscillations have been taken into account in the model AJCI ("Atome dans le jellium de charge impos\'ee") proposed by F. Perrot \cite{PerrotReport}. In this model again, similarly as in \textsc{Inferno}, all wave functions are normalized in the whole space. In addition to this, following ideas from earlier studies in the condensed matter on impurities in jellium, F. Perrot has abandoned the approximation of the neutrality of the WS sphere. He has assumed only that the total SCF potential be local due to screening. This assumption is equivalent to the neutrality of all charges in the whole space. A serious problem which appears is how to determine the jellium electron density since the model in principle allows one to find a variational solution in the space of four parameters : $Z$, $n_{i}$, $T$, and the average number of free electron per atom $Z^*$. In such a way the AJCI model requires an additional equation allowing one to determine $Z^*$. F. Perrot has proposed to use $Z^*$ from the TF ion-in-cell model i.e. from outside of the AJCI variational scheme. Due to this additional condition the AJCI model does not respect the virial theorem.

First satisfactory answer to the question how to formulate a fully variational approach to the quantum atom in jellium has been proposed in \cite{Blenski07a,Blenski07b,Piron09}. The difference with respect to the previous approaches \cite{Liberman79,Wilson06,PerrotReport} is the use of the cluster expansion for the two quantities : the plasma free energy per ion and the total number of electrons per ion. In both cases the zero order and the first order terms are retained from this expansion. As is shown in the cited references (see also Sec.~2), this systematic approach allows one to construct the variational system of equations respecting the virial theorem and to get a simple expression for the electronic pressure analogous to the one from \cite{Feynman49}. As concerns the SCF equations for the bound and free electron density, our approach is similar to the one of the AJCI model. The difference with respect to this model is the presence of a supplementary equation allowing one to determine $Z^*$. This equation stems from the variational relation requiring that the total free energy per ion be stationary with respect to the average electron density $n_{0}$. This is the only variational equation in which both zero order and the first order terms of the cluster expansion are accounted for. This equation has been absent in the AJCI model since in this model as well as in the \textsc{Inferno} model the zero order terms in the cluster expansion have been disregarded. In the case of the \textsc{Inferno} model the requirement that the free energy be stationary with respect to $n_{0}$ is replaced by the requirement that the WS 
sphere be neutral. In the case of the present model the WS sphere is in principle not neutral. Let us mention also that in the case of the quasi-classical Thomas-Fermi expressions for the free energy and total number of electrons per ion, the approach from \cite{Blenski07a,Blenski07b} leads correctly to the TF ion-in-cell model of Feynman \ea \cite{Feynman49}.

This paper presents the main features of our VAAQP code which is based on the theory presented in Refs. \cite{Blenski07a,Blenski07b} and reports some preliminary results from this code. A first exploratory study (see \cite{Piron09}) has already shown the feasibility of calculations with VAAQP in the Warm Dense Matter (WDM) regime. It appeared that significant effects of the variational treatment could be expected in the WDM regime. In this paper, we focus on the thermodynamic coherence of the results obtained with the code and on a first comparison with an \textsc{Inferno}-type model. In Sec.~2 the main features of the model will be summarized. The code will be described and some of the main results of the previous study will be briefly recalled in Sec.~3. In Sec.~4 we will present the dependence of the electronic pressure versus density at constant temperature equal to $10$ $eV$ in case of an aluminium plasma in order to prove the thermodynamic consistency of our model. Sec.~5 is devoted to a first comparison with an \textsc{Inferno}-type model.


\section{Variational theory of average-atom in jellium}
In the cluster expansion the thermodynamic averages are expanded in the numbers of ions taken into account. As said before we first focus our attention of the free energy. In the zero order ("no ions") all ions are smeared out into a neutralizing background and the free energy per unit volume $f_{0}(n_{0};T)$ is that of a homogeneous electron gas at the temperature T and of an electron density $n_{0}$ which at this moment is considered as unknown :
\begin{equation}
\label{eq_def_Z}
n_0=n_i Z^*
\end{equation}
$n_0$ is also related in the usual way to the unknown chemical potential $\mu_0$.

In the first order we consider one ion while all other ions participate in the surrounding jellium. The central ion will be interacting with the jellium. The cluster expansion gives :
\begin{equation}
\langle f\rangle_1=n_i\int\,d^3r\left\{f_1^{ion+jellium}\left(X;n_i,Z,T;\vec{r}\right)-f_0\left(n_0,T\right)\right\}
\end{equation}
The subtraction of the zero order free energy in the integral assures the convergence of the term.
In this way we define the following expression for the free energy per ion :
\begin{equation}
F_0=F_0\left(n_0,n_i,T\right)\equiv\frac{f_0\left(n_0,T\right)}{n_i}
\end{equation}
\begin{equation}
F\left(X;n_i,Z,T\right)=F_0+\Delta F_1
\end{equation}
\begin{equation}
\Delta F_1=\int\,d^3r\left\{f_1^{ion+jellium}\left(X;n_i,Z,T;\vec{r}\right)-f_0\left(n_0,T\right)\right\}
\end{equation}
The first order contribution to the free energy per ion in our model consists of the three terms :
\begin{equation}
\Delta F_1=\Delta F_1^{0}+\Delta F_1^{el}+\Delta F_1^{xc}
\end{equation}

$\Delta F_1^{0}$ is the free energy of independent electrons (kinetic-entropic term) in a trial potential $V(r)$ which we assume to be localized.

$\Delta F_1^{el}$ is the electrostatic interaction term taking into account all charge densities. The electron density is no more homogeneous and depends on the position $\vec{r}$. For large values of the radius $r$ this density tends to the value of jellium electron density of the zero order $n_0$. The ion contribution to the charge density is due to the central nuclear point charge and to the charge density due to the non-central ions. We assume that this charge density is equal to $n_0$ beyond a cavity of radius $R$ and is equal to zero inside this cavity. In this way following the idea from Refs~\cite{Liberman79,PerrotReport} we introduce a cavity into which the non-central ion charge density does no enter. The only difference is that in our approach at this moment the cavity radius is still arbitrary.
In such a way we get for the electrostatic interaction term, in atomic units:
\begin{equation}\begin{split}
\Delta F_1^{el}=-&\int\,d^3r\Biggl\{ \left(n(r)-n_0\theta(r-R)\right)\Biggr.\cdot\\
&\Biggl.\left( \frac{Z}{r}-\frac{1}{2}\int\,d^3r'\left\{ \frac{n(r')-n_0\theta(r'-R)}{|\vec{r}-\vec{r'}|} \right\} \right) \Biggr\}
\end{split}\end{equation}

$\Delta F_1^{xc}$ is the exchange correlation term:

\begin{equation}
\Delta F_1^{xc}=\int\,d^3r\left\{ f_{xc}\left(n(\vec{r})\right) -f_{xc}\left(n_0\right) \right\}
\end{equation}

The total free energy per ion shall be minimized with respect to the trial potential $V(r)$ and the average free electron density $n_0$ taking into account the two following supplementary conditions:
\begin{itemize}
\item The total charge neutrality:
\begin{equation}
\label{eq_total_neutr}
Z+\int\,d^3r\left\{ \left(n(r)-n_0\theta(r-R)\right)\right\}=0
\end{equation}
\item The "ionization model" which we get from the two first terms of the cluster expansion
applied to the total number of electrons per ion :
\begin{equation}
\label{eq_electron_cluster}
Z n_i=n_0+n_i\int\,d^3r\left\{ n(r)-n_0\right\}+...
\end{equation}
\end{itemize}
The substitution of Eq.~\ref{eq_electron_cluster} into Eq.~\ref{eq_total_neutr} leads immediately to the conclusion that the cavity radius $R$ should be the WS radius :
$R=R_{WS}$.

The explicit minimization with respect to the trial potential $V(r)$ and the average free electron density $n_0$ is thus applied to the thermodynamic potential 
\begin{equation}\begin{split}
\Omega &\left(n_0,V(\vec{r});n_i,Z,T \right) =F\left(X;n_i,Z,T\right) \\
&-\gamma\left(Z+\int\,d^3r\left\{n(r)-n_0\theta(r-R_{WS})\right\}\right)
\end{split}\end{equation}
where $\gamma$ is a Lagrange multiplier.

This minimization procedure leads, in the Local Density Approximation (LDA) to the SCF equations:
\begin{equation}
V(r)=V_{el}(r)-V_{xc}\left(n(r)\right)+V_{xc}\left(n_0\right)
\end{equation}
\begin{equation}
\gamma=\mu_0+V_{xc}(n_0)
\end{equation}
where $V_{el}(r)$ is the electrostatic part of the potential:
\begin{equation}
V_{el}(r)=\frac{Z}{r}-\int\,d^3r'\left\{ \frac{n(r')-n_0\theta(r'-R)}{|\vec{r}-\vec{r'}|} \right\}
\end{equation}
and
\begin{equation}
V_{xc}(n)=\frac{\partial f_{xc}(n)}{\partial n}
\end{equation}

As said before, we get also a condition which allows one to determine $n_0=n_i Z^*$:
\begin{equation}
\label{eq_variationnelle}
\int\,d^3r\left\{V_{el}(\vec{r})\theta(r-R_{WS})\right\}=0
\end{equation}

Explicit analytical differentiation with respect to ion density $n_i$ of the thermodynamic potential at the equilibrium leads to the following formula for the electronic pressure:
\begin{equation}
\label{eq_pression}
P=n_i^2\left.\frac{\partial \Omega}{\partial n_i}\right|_{expl.}=-f_0+n_0\left(\mu_0+V_{xc}(n_0)+V_{el}(R_{WS})\right)
\end{equation}

\section{The VAAQP code}
\begin{figure}[t]
\centerline{\includegraphics[width=8cm]{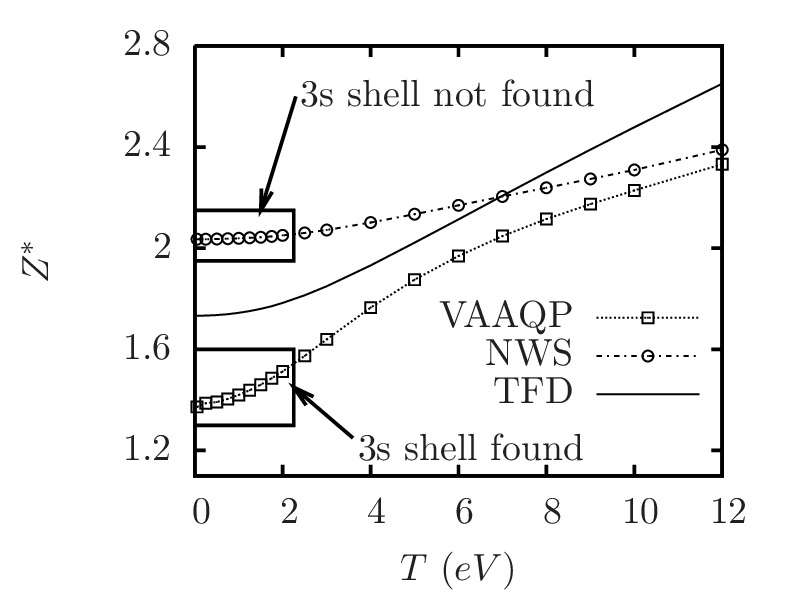}}
\caption{Comparison between mean ionizations obtained from variational calculation (VAAQP), from non-variational NWS calculation (NWS), and from TFD calculation (TFD).
\label{fig3}}
\end{figure}

The VAAQP code solves the equations of the model described above. In the present version first it finds solutions of the SCF equations in the four-parameter space: ($Z$, $T$, $n_i$, $Z^*$) i.e. it proceeds as if an equilibrium could exist for any chosen value of the mean ionization $Z^*$. Then it finds among these solutions the one that corresponds to the correct equilibrium i.e. such solution which fulfills Eq.~\ref{eq_variationnelle}.

This code can calculate the electron density and the Average Atom structure using semiclassical (TF), quantum-non-relativistic (Schroedinger) or quantum-relativistic (Dirac) formalism. Finite difference methods are being used to compute electron density as function of the radius $r$ up to an asymptotic radius at which the electron density is matched to its asymptotic form taken from the finite-temperature linear response theory \cite{Khanna76,Gouedard78}. The iteration scheme previously used by Jena and Singwi (see \cite{Jena78}) is used to obtain the self-consistency between the electron density and the potential $V(r)$.

At present, the code can use Iyetomi and Ichimaru finite-temperature exchange-correlation free energy term \cite{Iyetomi86} as well as Dirac exchange energy term \cite{Dirac30,KohnSham65a} but any other local-density exchange-correlation free energy term can also be used. In all the preliminary studies that we present in this paper, Dirac exchange term was chosen in order to provide direct comparisons with the Thomas-Fermi-Dirac (TFD) model. In the temperature regime of interest in these studies, the differences between results obtained using Iyetomi exchange-correlation term and using Dirac exchange term appeared to be relatively small (the effect of finite temperatures seems to be small).

As stated before this code is primarily designed to calculate solutions of our variational model (which will be denoted by VAAQP). It has however other options. One of these allows one to perform the atom-in-jellium calculation replacing the variational equation by the neutrality of the WS sphere (the corresponding calculation will be denoted by NWS). Another option allows one to find \textsc{Inferno}-type solutions in which all calculations are limited to the neutral WS sphere. In this way, we can compare results from different approaches using exactly the same numerical procedures for the calculation of the electron density and other structure quantities.

The VAAQP code was first presented in \cite{Piron09} and it has been shown there using the example of the isochoric EOS curve corresponding to Al at $2.7$ $g.cm^{-3}$ matter density that variational treatment could lead to substantial differences of the calculated thermodynamic quantities with respect to results from non-variational models. These differences could be of the same order as the effect of the exchange-correlation or the effect of the quantum treatment of the electrons. They appeared to be especially important in the case of the WDM regime which is characterized in our model by weakly damped Friedel oscillations. As an example, we recall FIG.~\ref{fig3} which presents the mean ionization as defined in Eq.~\ref{eq_def_Z} along the Al $2.7$ $g.cm^{-3}$ isochoric curve for three models : VAAQP, NWS and TFD. Differences can be of the order of 30\% and it can be seen that below $2.25$ $eV$ temperature, even the shell structure of the Average Atom is affected.

\section{Direct verification of the variational principle along the aluminium 10eV isotherm curve}

\begin{figure}[t]
\centerline{\includegraphics[width=8cm]{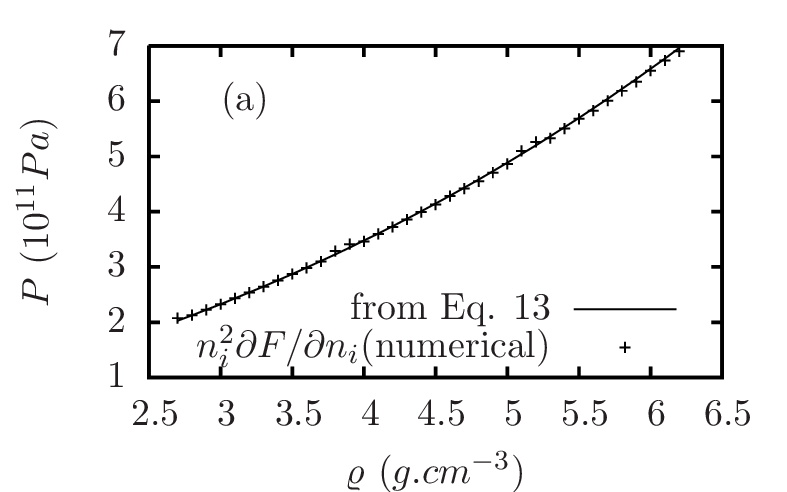}}
\centerline{\includegraphics[width=8cm]{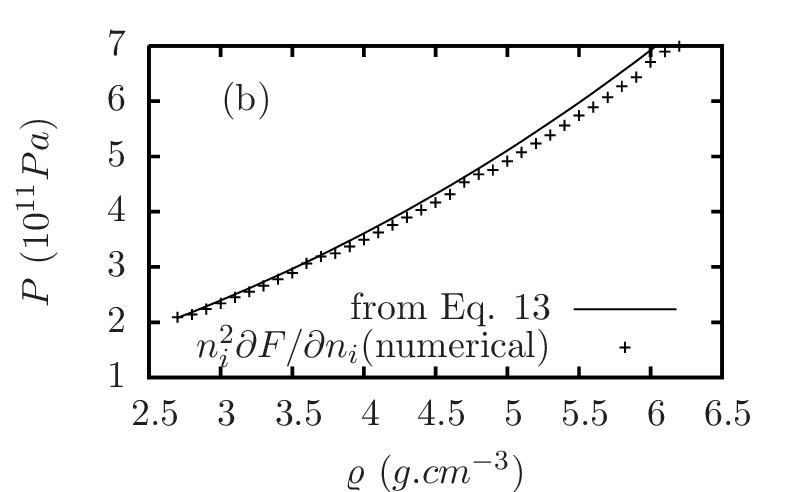}}
\caption{Comparison between electronic pressures obtained from Eq.~\ref{eq_pression} and by numerical differentiation of the free energy along the $10$ $eV$ isotherm curve for variational calculations (a) and non-variational NWS calculations (b).
\label{fig5-6}}
\end{figure}

The electronic pressure formula from Eq.~\ref{eq_pression} is obtained by differentiation of the equilibrium free energy with respect to the ion density $n_i$ taking into account the conditions stated in Eqs.~\ref{eq_electron_cluster}, \ref{eq_total_neutr}. As it is well known (see for instance \cite{Blenski07b}), this is equivalent to the explicit analytical differentiation of the grand potential $\Omega$ with respect to the ion density. Each solution of a calculation with the variational option of VAAQP corresponds to the minimum free energy respecting Eqs.~\ref{eq_electron_cluster}, \ref{eq_total_neutr} i.e. the explicit minimum of $\Omega$. Therefore, direct numerical differentiation of the free energy obtained with the variational option of VAAQP along an isotherm curve will also give the electronic pressure. In the case of a non-variational calculation such as with the NWS option of VAAQP, the grand grand potential $\Omega$ is not minimized. Calculations of the electronic pressure from Eq.~\ref{eq_pression} and from numerical differentiation of the free energy should thus lead to values that are different. In case of a non-variational approach none of these constitutes a justified way to obtain the electronic pressure as it is defined in thermodynamics. Nonetheless, numerical differentiation of the free energy is often used as a kind of a "mechanical" definition of the electronic pressure in some existing codes.

In order to check the thermodynamic consistency of VAAQP variational option we can compare electronic pressures calculated from Eq.~\ref{eq_pression} to the one obtained by numerical differentiation of the equilibrium free energy. Checking the agreement between these two quantities is not only a proof of the variational character of the model but can also be viewed as a very severe test of the performance of the code. We recall that such thermodynamic consistency has never been possible with other existing quantum Average Atom models.

\begin{figure}[t]
\centerline{\includegraphics[width=8cm]{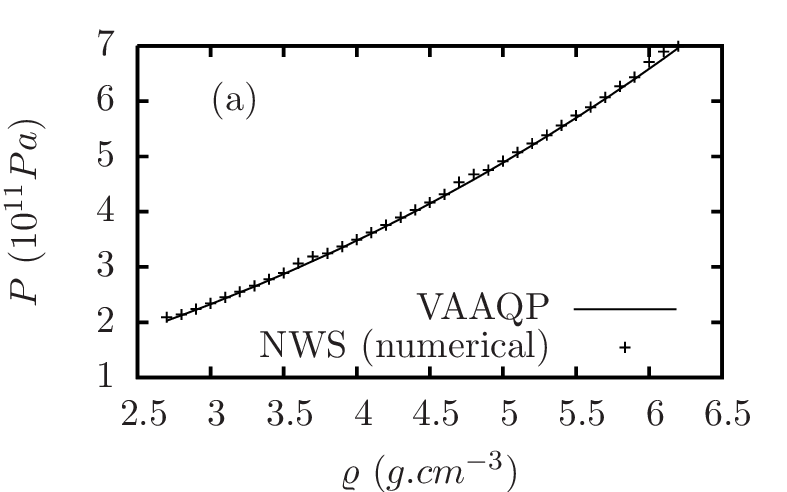}}
\centerline{\includegraphics[width=8cm]{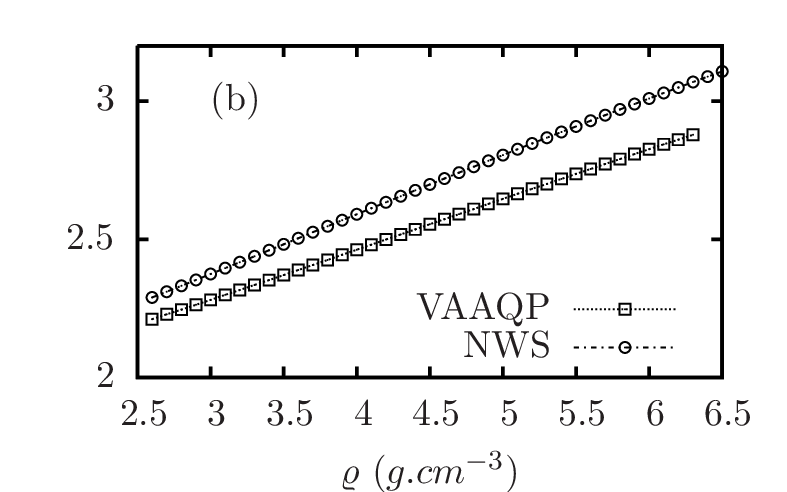}}
\caption{Comparison between electronic pressures (a) and mean ionizations (b) obtained along the $10$ $eV$ isotherm curve for variational calculations (VAAQP) and non-variational NWS calculations (NWS). Pressures presented on (a) for NWS calculations stem from the numerical differentiation of the free energy.
\label{fig7-8}}
\end{figure}

Comparisons between electronic pressures calculated from Eq.~\ref{eq_pression} and from numerical differentiation of the free energy along the Al $10$ $eV$ isotherm curve is displayed on FIG.~\ref{fig5-6}. The case of VAAQP variational option is addressed on FIG.~\ref{fig5-6}a whereas the case of the non-variational NWS option is presented on FIG.~\ref{fig5-6}b. As expected, the agreement is found in the case of the variational option whereas in the case of the non-variational option a disagreement can be observed. It is important to note that these differences tend to increase in the WDM regime. As it has been already noted in \cite{Piron09}, rather important effects of variational treatment are to be expected in such regimes where the Friedel oscillations decay on a scale much larger than the WS radius.

It is also interesting to compare the electronic pressures obtained from VAAQP variational option to those calculated by numerical differentiation of the free energy in the case of the non-variational NWS option. As it can be seen on FIG.~\ref{fig7-8}a, these two electronic pressures obtained along the Al $10$ $eV$ isotherm curve are rather close. Nevertheless mean ionizations obtained from these two options can differ significantly, of the order of 10\% (see FIG.~\ref{fig7-8}b). It means that average electron densities obtained from the two models can differ significantly despite the fact that both option lead to electronic pressures that are relatively close. It seems therefore that comparisons between models should thus be made not only as concerns the EOS curves but also by comparison of other atomic physics properties involving the atomic structure such, for instance, as electrical conductivity.

\begin{figure}[t]
\centerline{\includegraphics[width=8cm]{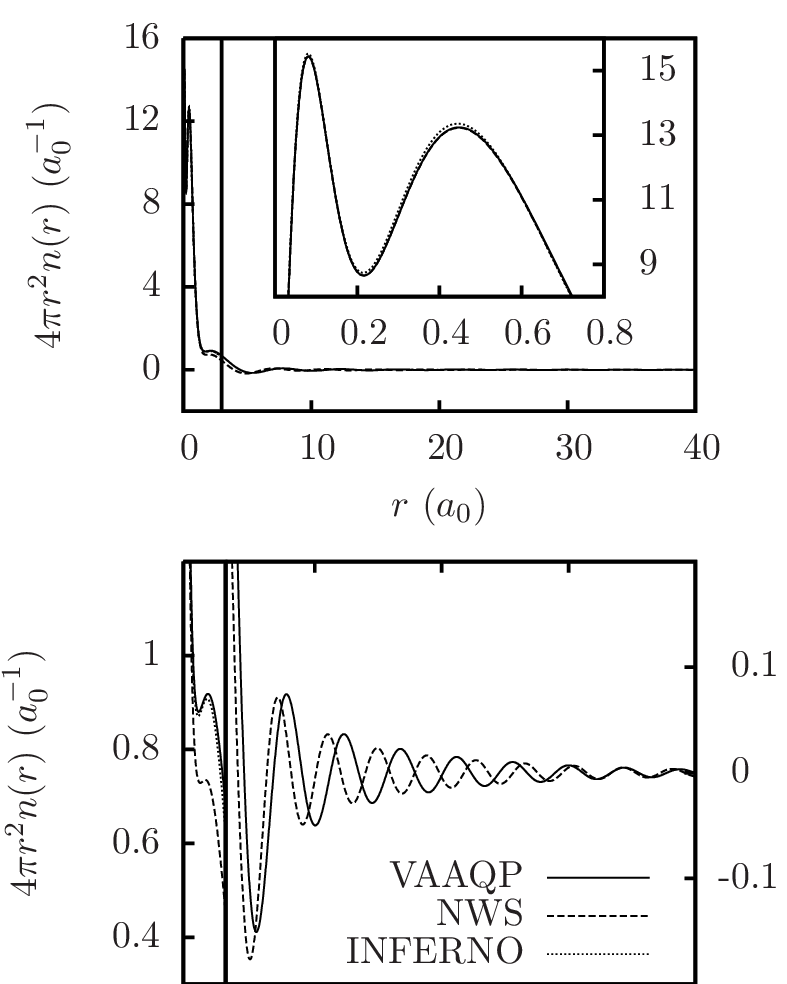}}
\caption{Comparison between electron density obtained for Al at $0.25$ $eV$ temperature and $2.7$ $g.cm^{-3}$ matter density from a variational calculation (VAAQP), a non-variational NWS calculations (NWS), and an \textsc{Inferno}-type calculation (INFERNO). Global plot of the electron density is presented, as well as enlarged views of the shell structure, Friedel oscillations and vicinity of the WS radius.
\label{fig9}}
\end{figure}
\section{First comparison with \textsc{Inferno}-type model}
As a first step to comparison with \textsc{Inferno}-type models, we present in FIG.~\ref{fig9} a comparison of the electron density obtained for the three models corresponding to the VAAQP variational option, the non-variational NWS option, and the \textsc{Inferno} option. 

Again, it is worth to note that results from the variational option and from the non-variational NWS option differ strongly in the region dominated by Friedel oscillations. It is interesting to see that although \textsc{Inferno} model is based on the neutrality of the WS sphere, in the vicinity of the WS radius the \textsc{Inferno}-type electron density seems to be much closer to the variational solution than to the electron density from the NWS option. The effect of the \textsc{Inferno}-type treatment seems to be important for the shell structure which appears to be different both from the ones of the VAAQP option and of the NWS option.

\section{Conclusion}
In this paper, results of a new code based on a fully variational model of dense plasma at thermodynamical equilibrium are reported. This model takes into account the behavior of the self-consistent density and potential beyond the WS radius and allows one to obtain thermodynamically coherent value of the mean ionization. 

Thermodynamic consistency of the results obtained with the code is shown on the example of the aluminium $10$ $eV$ isotherm curve. The results from this case show also that different models can lead to similar electronic pressures while providing significantly different average electron densities i.e. mean ionizations.

The work on comparisons between thermodynamical quantities and atomic physics properties obtained from the variational VAAQP option and from the \textsc{Inferno}-type models is still in progress. As a first step, comparison of the electron density obtained with three models including an \textsc{Inferno}-type model is given.

\section{Acknowledgments}
The authors would like to thank Brian Wilson and Philip Sterne for fruitful discussions on the \textsc{Purgatorio} code.

They would also like to thank Philippe Arnault, Christopher Bowen, Alain Decoster, G\'erard Dejonghe, G\'erald Faussurier, Jean-Christophe Pain, Fran\c cois Perrot, David Bailey, Carlos Iglesias, and Stephen Libby for helpful discussions.

This work has been supported by the European Communities under the contract of association between EURATOM and CEA within the framework of the European Fusion Program. The views and opinions expressed herein do not necessarily reflect those of the European Commission.


\end{document}